%% file: main.tex
\documentclass[a4paper, amsfonts, amssymb, amsmath, reprint, showkeys, nofootinbib, twoside, preprintnumbers]{revtex4-1}
\usepackage[english]{babel}
\usepackage[utf8]{inputenc}
\usepackage[colorinlistoftodos, color=green!40, prependcaption]{todonotes}

\input{preamble}
\usepackage[pdftex, pdftitle={Article}, pdfauthor={Author}]{hyperref} 
\begin{document}
\title{Intrinsic Quality Factor Extraction of Multi-Port Cavity with Arbitrary Coupling}

\author{D. Frolov}
    \email[Correspondence email address: ]{dfrolov@fnal.gov}
    \affiliation{Fermi National Accelerator Laboratory, Batavia, IL 60510, United States}
    
\date{\today} 
\preprint{FERMILAB-PUB-20-197-TD}

\begin{abstract}
We derived S-parameter based expressions for the intrinsic quality factor of an arbitrary coupled multiport microwave cavity and non-ideal test fixture.  Practical accuracy limitations of the obtained expressions specifically for superconducting accelerator cavities were evaluated both analytically and with the simulation software. The resulting formulas can be used to extract intrinsic quality factor of normal conducting and superconducting cavities directly from calibrated S-parameter measurements.
\end{abstract}


\maketitle
\section{Introduction}
In superconducting RF accelerator field (SRF) intrinsic quality factor $Q_0$ of a cavity is the key parameter. Practical and theoretical developments in material science, aimed to improve cavity performance through surface physics understanding completely rely on $Q_0$ measurement, therefore it is critically important to ensure measurement accuracy and repeatability. 

Determination of the intrinsic quality factor and accelerating gradient is usually based on the analysis of the RF power that is reflected from the cavity, dissipated in the cavity, and transmitted through it. For example, Fermilab and JLab are using procedure \cite{powers} for $Q_0$ measurement in vertical test stand (VTS). This procedure is based on methods of scalar network analysis (based on power measurement), however these methods are approximate by definition because exact wave distribution in the RF structures can be described only with complex numbers (vectors). Therefore, such scalar methods can contain significant errors because features of the RF networks with distributed parameters are not considered. Accuracy of the procedure \cite{powers} was studied in \cite{Melnychuk}, however vector errors were not taken into account as well as some mismatch errors. 

More accurate methods of vector network analysis (based on magnitude and phase measurements) can be used for $Q_0$ measurement in VTS. Several ideas of such measurements with elements of vector correction have been proposed earlier. In \cite{Holzbauer}, \cite{Holzbauer2} authors used heuristic approach to demonstrate errors caused by the mismatches in the VTS RF system and their dependence on phase shifts in the cavity input transmission line. Much earlier at CERN \cite{CERN} contribution of test fixture non-idealities to $Q_0$ error in scalar form was demonstrated. Recently in \cite{Holzbauer3} authors successfully used variable delay line installed between cavity and incident power source to characterize vector errors, a very well-known method used in the six-port reflectometers \cite{six-port1}. And in \cite{Goryashko} authors used another method - cavity in a self-excited loop connected to a commercial vector network analyzer (VNA) to obtain vector data. 

On the other hand, these previous studies were done for single port or two-port cavities, do not have complete RF system models developed and only partially use the advantages of vector measurements: effects of field probe load mismatch are not studied, multiport cavities are not studied, there are no general equations allowing to extract arbitrary coupled cavity $Q_0$ only from the S-parameters.

This paper is focused on the derivation of such practically useful equations and their verification. In Section II we present the initial mathematical foundations required for the analysis and derive general coupling equation for a cavity with any number of ports and its solution. Then using this result, in Section III we show how to extract cavity quality factor purely from its S-parameters in form of the exact solution and then we formalize the two types of approximations that are required because of practical limitations of the superconducting accelerator cavities measurements. In Section IV we analyze accuracy of the obtained two types of approximations by comparing them numerically with the exact solution. In Section V the results of simulations using the obtained equations Keysight Genesys RF/microwave circuit synthesis and simulation software are presented. The paper ends with the conclusions in Section VI.  

The results obtained in this paper can be used to extract intrinsic quality factor of normal- and superconducting cavities directly from S-parameter measurements. Specific calibration procedures required to obtain S-parameters of superconducting cavities exist but are not described in this paper. We, however, show several fundamental limitations of such S-parameter measurements.

\section{Initial Equations}
\begin{figure*}[!htb]
\includegraphics{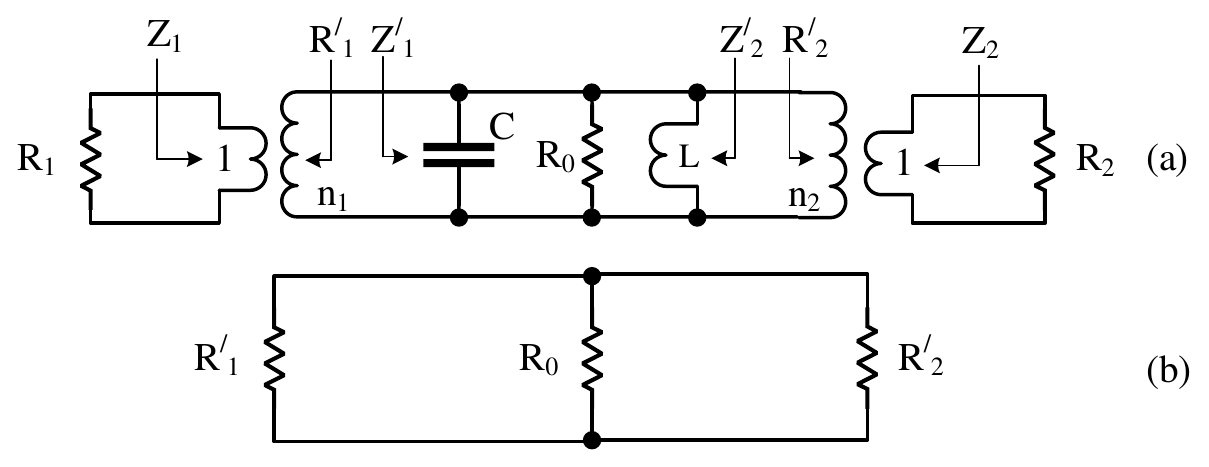}
\caption{\label{fig:model} (a)- Lumped element two-port cavity model. $R_1, R_2$- input and output test port impedance, $Z_1, Z_2$- input and output cavity port impedance, same symbols denoted by $^/$ are cavity port and test port impedances as seen by the cavity itself as a RF source. $R_0, C, L$ - cavity equivalent shunt resistance, capacitance, and inductance. (b)- equivalent circuit of model.}
\end{figure*}

In many previous works equations for cavity coupling and quality factors are derived based on lumped-element cavity model. These known derivations often assume perfect test port coupling and are limited to a single port or a two-port cavity. In this section we will derive port couplings and $Q_0$ of a \textit{multiport} cavity using lumped-element cavity model (a) and its impedance equivalent (b) in fig.~\ref{fig:model}.

Model (a) contains RF source impedance $R_1$, input coupler in form of an equivalent transformer with ratio~$1:n_1$, cavity shunt resistance $R_0$, equivalent L-C circuit, output coupler (field probe) with ratio~$n_2:1$ and field probe load impedance $R_2$.
  
The intrinsic $Q_0$ and loaded $Q_L$ quality factors of parallel L-C circuit are defined as (\ref{eq:Qs}).
	\begin{equation}\label{eq:Qs} 
	   \frac{1}{Q_0} = \frac{1}{R_0} \sqrt{\frac{L}{C}} ,\quad 
	   \frac{1}{Q_L} = \Big| \frac{1}{R_L} \Big| \sqrt{\frac{L}{C}}. \\ 
     \end{equation}
From which $Q_0$ can be expressed as (\ref{eq:QL}).
    \begin{equation}\label{eq:QL} 
    Q_0 = Q_L \Big| \frac{R_0}{R_L} \Big|. 
     \end{equation}          
Load impedance $R_L$ can be found from equivalent circuit of the model fig.~\ref{fig:model}~(b) as (\ref{eq:ZL}).           
    \begin{equation}\label{eq:ZL} 
    \frac{1}{R_L} = \frac{1}{R_0} + \frac{1}{R_1^/} + \frac{1}{R_2^/}. 
     \end{equation}     
By substituting (\ref{eq:ZL}) into (\ref{eq:QL}) and replacing impedance ratios with (\ref{eq:betasg}) intrinsic quality factor $Q_0$ can be written in form (\ref{eq:Q0}).
	
	\begin{equation}\label{eq:betasg} 
	   \beta_1 \equiv R_0/R_1^/ ,\quad 
	   \beta_2 \equiv R_0/R_2^/ . 
    \end{equation}

    \begin{equation}\label{eq:Q0} 
    Q_0 =Q_L | 1 + \beta_1 + \beta_2 |. 
    \end{equation} 
By continuing this reasoning one can easily show fairness of expression (\ref{eq:Q0_general}) for a cavity with $n$ number of ports.  
    \begin{equation}\label{eq:Q0_general} 
    Q_0 =Q_L \Big| 1 + \sum_{n}  \beta_n \Big| . 
    \end{equation}  

The loaded quality factor $Q_L$ is trivial and can be measured from the 3~dB bandwidth of cavity or from decay fit. In the previous studies \cite{Holzbauer} authors considered systematic errors in $Q_L$ measurement as deviation of $Q_L$ measured with mismatched load $R_1$ from the ideal case with matched load.  However, we find that more natural approach coming from the definition of the loaded quality factor is to consider that $Q_L$ is always measured correctly, because it is determined by the actual \textit{load}, that is not necessary matched. Therefore $Q_L$ error doesn't depend on RF system mismatches, but errors exist in $\beta_n$ measurements, so the goal is to find general definitions of $\beta_n$ for arbitrary $Z_n$ and $R_n$. 

We will start with two-port cavity and then expand results for multiport cavity. Here the notation is used where coefficients $\beta_1^/$ of the cavity input port, $\beta_2^/$ of output port, $\alpha_1$ of the RF source, $\alpha_2$ of field probe load express their corresponding couplings to the transmission line with impedance $Z_0$. For exact calculations these coupling coefficients \textit{must} be complex numbers, since impedances $Z_1$, $Z_2$ and $R_1$, $R_2$ may not be purely resistive in general. Those skilled in the art may notice that in some SRF papers \cite{powers}, \cite{Melnychuk} absolute value $|\beta_1^/|$ is also known as $\beta^*$ - the overall input coupling coefficient \cite{mende}. For model fig.~\ref{fig:model}(a) the following equations are valid:

	\begin{equation}\label{eq:impedance}
		R_1^/ = R_1 \cdot n_1^2, \quad
		R_2^/ = R_2 \cdot n_2^2, 
    \end{equation}    
	\begin{equation}\label{eq:transformer_ratio}
		n_1^2 = Z_1^/ / Z_1, \quad
		n_2^2 = Z_2^/ / Z_2, 
    \end{equation}   
	\begin{equation}\label{eq:betaStar}   
		\beta_1^/ \equiv Z_1/Z_0, \quad
		\beta_2^/ \equiv Z_2/Z_0, 
    \end{equation}
	\begin{equation}\label{eq:alpha} 
		\alpha_1 \equiv R_1/Z_0, \quad
		\alpha_2 \equiv R_2/Z_0.  
    \end{equation}      
By substituting expressions (\ref{eq:transformer_ratio}), (\ref{eq:betaStar}), (\ref{eq:alpha}) into (\ref{eq:impedance}) one will find ratios (\ref{eq:r1r2}).

	\begin{equation}\label{eq:r1r2} 
		\frac{R_1^/}{Z_1^/} = \frac{\alpha_1}{\beta_1^/} , \quad
		\frac{R_2^/}{Z_2^/} = \frac{\alpha_2}{\beta_2^/} .  
    \end{equation}
From equivalent circuit in fig.~\ref{fig:model}(a) $Z_1^/$ and $Z_2^/$ can be found in form (\ref{eq:z1z2}). By substituting (\ref{eq:z1z2}) into (\ref{eq:r1r2}) and replacing $R_1^/$ and $R_2^/$ with $R_0/ \beta_1$, $R_0/\beta_2$ using (\ref{eq:betasg}) we finally find pair of linear equations (\ref{eq:beta_equations_pair}).   

	\begin{equation}\label{eq:z1z2} 
		\frac{1}{Z_1^/} = \frac{1}{R_0}+ \frac{1}{R_2^/}, \quad
		\frac{1}{Z_2^/} = \frac{1}{R_0}+ \frac{1}{R_1^/}.  
    \end{equation}

	\begin{equation}\label{eq:beta_equations_pair}
		\frac{\alpha_1}{\beta_1^/} \beta_1 - \beta_2  = 1, \quad
		\frac{\alpha_2}{\beta_2^/} \beta_2 -\beta_1  = 1. 
    \end{equation}
Solution to this system can be easily found in form (\ref{eq:beta_solution_pair}) where $\gamma_n = \alpha_n / \beta_n^/$.

	\begin{equation}\label{eq:beta_solution_pair}
		\beta_1 = \frac{\gamma_2 + 1}{\gamma_1 \gamma_2 - 1}, \quad
		\beta_2 = \frac{\gamma_1 + 1}{\gamma_1 \gamma_2 - 1}. 
    \end{equation}

Addition of more ports to the cavity is equivalent to addition of more resistors in parallel to the circuit in fig.~\ref{fig:model}(b), more $1/R^/$ fractions to equation (\ref{eq:ZL}), and hence increasing total number of the equations. For three ports (\ref{eq:beta_equations_pair}) will turn into three equations: 
\begin{align*}     
& \frac{\alpha_1}{\beta_1^/} \beta_1 - \beta_2 - \beta_3 =1, \\   
& \frac{\alpha_2}{\beta_2^/} \beta_2 - \beta_1 - \beta_3 =1, \\  
& \frac{\alpha_3}{\beta_3^/} \beta_3 - \beta_1 - \beta_2 =1.  
\end{align*}

By continuing the above reasoning one can show that for N-port cavity (\ref{eq:beta_equations_pair}) will turn into system of N equations with equation number $n$ in form (\ref{eq:quation_summ}).  

    \begin{equation}\label{eq:quation_summ} 
    \gamma_n \beta_n -  \sum_{i=1}^N \beta_i + \beta_n  = 1
    \end{equation}  
Therefore general solution of such system (\ref{eq:quation_summ}) for the coupling $\beta_n$ for an N-port cavity with arbitrary $Z_n$ and $R_n$ can be written in the matrix form (\ref{eq:quation_summ_gen_solution}).  

    \begin{equation}\label{eq:quation_summ_gen_solution} 
     \begin{pmatrix}
         \beta_1 \\
         \beta_2 \\
         \vdots \\
         \beta_N    
    \end{pmatrix} 
    =  
    \begin{pmatrix}
         \gamma_1  & -1       & \hdots     & -1 \\   
         -1        & \gamma_2 & \hdots     & -1 \\   
         \vdots    & \vdots   & \ddots     & \vdots \\
         -1        & -1       & \hdots     & \gamma_N     
    \end{pmatrix}^{-1} 
    \begin{pmatrix}
         1 \\
         1 \\
         \vdots \\
         1    
    \end{pmatrix}    
    \end{equation}
For the one-port cavity $\beta_1=\beta_1^/ / \alpha$ is quite simple and in agreement with previous work \cite{CERN}, however for the two-port cavity $\beta_1 = \frac{\beta_1^/(\alpha_2 + \beta_2^/)}{\alpha_1 \alpha_2 - \beta_1^/ \beta_2^/}$ is more complicated and complexity increases with number of ports. 

General solution (\ref{eq:quation_summ_gen_solution}), expression (\ref{eq:Q0_general}) and definition of $\gamma_n$ lead to a conclusion that for a cavity with any number of ports information about loaded quality factor $Q_L$, \textit{reflection} coefficients from the cavity ports \textit{and} test ports is \textit{sufficient} to determine $Q_0$.      

\section{$Q_0$ Extraction from S-parameters}
\subsection{Exact solution}

We express coupling to the transmission line with impedance $Z_0$ using complex reflection coefficients of the cavity ports $\Gamma_n$ corresponding to impedances $Z_1 \hdots Z_n$ and test ports $L_n$ \footnote{$L_n$ should not be confused with inductance.} corresponding to impedances $R_1 \hdots R_n$ in fig.~\ref{fig:model}(a). So coefficients (\ref{eq:betaStar}, \ref{eq:alpha}) can be reformulated in form (\ref{eq:Gamma}).

    \begin{equation}\label{eq:Gamma} 
    \begin{split}
        & \alpha_n = 1+L_n /(1-L_n), \\
        & \beta_n^/ =  1+\Gamma_n /(1-\Gamma_n ). \\
    \end{split}
    \end{equation}     
Hence, coupling ratio $\gamma_n$ in the solution (\ref{eq:quation_summ_gen_solution}) can be expressed in form  (\ref{eq:small_gamma_reflections}).
 
     \begin{equation}\label{eq:small_gamma_reflections} 
    \gamma_n = \frac{(1+L_n)(1 - \Gamma_n)}{(1-L_n)(1+\Gamma_n)}. 
    \end{equation} 
Using this notation, we find how $Q_0$ of a two-port cavity can be extracted purely form S-parameters. 

\begin{figure}[!htb]
\includegraphics{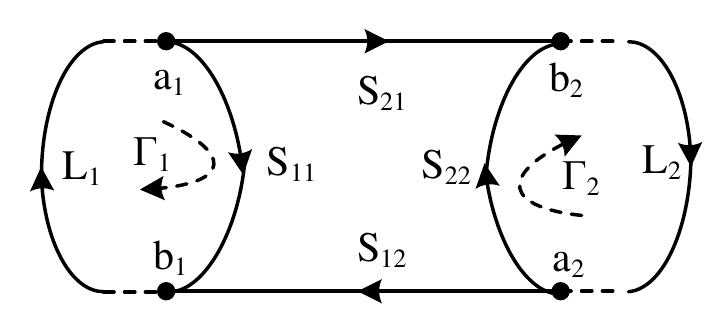}
\caption{\label{fig:graph} Flow graph of two-port cavity with source reflection $L_1$ and field probe load reflection $L_2$  }
\end{figure}

Input reflection coefficient $\Gamma_{1}$ of a two-port network terminated with load with reflection $L_2$ can be written as (\ref{eq:input_gamma1}) by applying Mason's rule \cite{mason_rule} to flow graph in fig.~\ref{fig:graph}.

    \begin{equation}\label{eq:input_gamma1} 
    \Gamma_{1} = S_{11} + \frac{S_{21} S_{12} L_2}{1-S_{22} L_2}
    \end{equation}   
By analogy, reverse reflection coefficient $\Gamma_{2}$ can be written as (\ref{eq:input_gamma2}). 

    \begin{equation}\label{eq:input_gamma2} 
    \Gamma_{2} = S_{22} + \frac{S_{12} S_{21} L_1}{1-S_{11} L_1}
    \end{equation}

Hence, by substituting (\ref{eq:input_gamma1}), (\ref{eq:input_gamma2}) into (\ref{eq:small_gamma_reflections}) and using (\ref{eq:beta_solution_pair}), (\ref{eq:Q0_general}), we find equation for $Q_0$ of a two-port cavity extracted purely from the cavity and test ports' S-parameters (\ref{eq:q0_sparam}). 

    \begin{equation}\label{eq:q0_sparam}
    \begin{split} 
    & Q_0 = Q_L | 1 + \beta_1 + \beta_2 | , \\
    & \beta_1 = \frac{(L_1-1)(S_{11} - L_2 S_{22} - B_2 + C_2 +1)}{S_{11} - L_2 - L_1 + S_{22} + D }  , \\
    & \beta_2 = \frac{(L_2-1)(S_{22} - L_1 S_{11} - B_1 + C_1 +1)}{S_{11} - L_2 - L_1 + S_{22} + D }  , \\
    &\text{where:} \\
    & B_n = L_n S_{11} S_{22}, \\
    & C_n = L_n S_{12} S_{21}, \\
    & D = L_1 L_2 S_{11} + L_1 L_2 S_{22} - B_1 + C_1  - B_2  + C_2. \\
    \end{split}
    \end{equation}   

Equation (\ref{eq:q0_sparam}) can be applied to measure $Q_0$ directly with the VNA, coefficients $L_1$ and $L_2$ will correspond to ESF and ELF complex error factors according to the 12-term error model \cite{dunsmore}, \cite{hiebel}, \cite{marks}, \cite{solt_original} and can be extracted from the VNA calibration data or measured directly by connecting test ports of the VNA one to one. 

    It can be seen, that exact $Q_0$ solution (\ref{eq:q0_sparam}) for 2-port cavity is quite bulky, and becomes significantly complicated if cavity has more ports, however using general definitions (\ref{eq:quation_summ_gen_solution}), (\ref{eq:small_gamma_reflections}) and Mason's rule from the flow-graph theory one can find such exact solution for cavity with any amount of ports and arbitrary coupling, based only on the information about the S-parameters, which can be directly measured with a multi-port VNA.   
    
    If the VNA has high quality test cables and fixtures, we can approximate $L_1 \approx 0$, $L_2 \approx 0$ and then (\ref{eq:q0_sparam}) will reduce to (\ref{eq:q0_sparam_reduced}).   

    \begin{equation}\label{eq:q0_sparam_reduced} 
    Q_0 \approx Q_L \Bigg| \frac{-2}{S_{11}+S_{22}}  \Bigg| 
    \end{equation}    

The above found equations can be used both for room temperature and superconducting cavities. Our experience with ultra-high quality superconducting cavities \cite{quantum_sasha} with $Q_0$ approaching $10^{12}$ have shown that inside dilution refrigerator, for example, cavity detuning due to microphonics is negligible, so the phase locked loop is not required and the S-parameters can be measured directly by moving calibration standards into cryogenic environment as it has been demonstrated in \cite{zoya}. Therefore, using the above obtained equations can be extremely useful for material characterization and quantum information applications.

\subsection{First order approximation}
Accelerator cavities have specific features that significantly complicate usage of (\ref{eq:q0_sparam}) for calculations. To achieve operating accelerating gradients of tens of megavolts per meter such cavities are tested at high levels of input power, where usually only one input port is near critical coupling ($\beta \approx 1$) to the RF source. The output ports are usually significantly undercoupled ($\beta \ll 1$) and used as electric field probes. Under such condition direct determination of cavity S-parameters becomes impossible because reflection measurements from the undercoupled ports demand for enormous power from the RF source to achieve high gradient of electric field in the cavity. Measurements at low power levels do not give accurate results for $Q_0$ because cavity parameters highly depend on gradient.
  
For many years different facilities used scalar power balance method to extract cavity ports' couplings and measure $Q_0$. This method requires critical coupling of input coupler which in practice is frequently violated and causes significant errors, as shown in previous studies \cite{Holzbauer}, \cite{Holzbauer2}. By using vector parameters instead, it is possible to achieve ultimate accuracy of cavity measurements in wide range of couplings. 

On the example of our two-port cavity model we analyze what fundamental limitations on $Q_0$ accuracy are imposed by the lack of information about $S_{22}$. For the reasons explained above, when testing accelerating cavity it is possible to physically measure only ratios $R_{11} = b_1/a_1$ and $T_{21} = b_2/a_1$ of incident $a_1$, transmitted $b_2$ and reflected $b_1$ waves of flow graph in fig.~\ref{fig:graph}. By applying Mason's rule to the flow graph, these ratios can be expressed as pair of linear equations (\ref{eq:reflection_transmission}).

    \begin{equation}\label{eq:reflection_transmission} 
    \begin{split} 
    & R_{11} = \frac{S_{21} L_2 S_{12} + S_{11} (1-L_2 S_{22}) }{1-(S_{21} L_2 S_{12} L_1 + S_{11} L_1 + L_2 S_{22}) + S_{11} L_1 L_2 S_{22}}, \\
    & T_{21} = \frac{S_{21} }{1-(S_{21} L_2 S_{12} L_1 + S_{11} L_1 + L_2 S_{22}) + S_{11} L_1 L_2 S_{22}}. \\
     \end{split} 
    \end{equation}

Such system (\ref{eq:reflection_transmission}) has four unknowns $S_{11}$, $S_{21}$, $S_{12}$, $S_{22}$ and cannot be solved in general. It is physically possible, however, to design RF system in such way, that reflection coefficient $L_2$ will be negligibly small and since $L_2 \ll S_{22}<1$  pair (\ref{eq:reflection_transmission}) will reduce to (\ref{eq:reflection_transmission_reduced}).   

    \begin{equation}\label{eq:reflection_transmission_reduced} 
    \begin{split} 
    \lim_{L_2 \to 0} R_{11} = \frac{S_{11} }{1- S_{11} L_1 }, \
    \lim_{L_2 \to 0} T_{21} = \frac{S_{21} }{1- S_{11} L_1 }. 
    \end{split}
    \end{equation}

From (\ref{eq:reflection_transmission_reduced}) one can find cavity $S_{11}$ and $S_{21}$ in form of first order approximation (\ref{eq:cav_s11_s21}) which will be valid only when $L_2 \to 0$. For N-port cavity, with $n=2 ... N$ under condition that $L_{n} \to 0$ it can be shown that transmission from the first port to $n$-port  is expressed as (\ref{eq:Tn}). 

    \begin{equation}\label{eq:cav_s11_s21} 
    \begin{split}
    S_{11} = \frac{R_{11}}{1+ L_1 R_{11}}, \
    S_{21} = \frac{T_{21}}{1+ L_1 R_{11}}.
    \end{split}
    \end{equation}     

    \begin{equation}\label{eq:Tn}
    S_{n,1} = \frac{T_{n,1}}{1+ L_1 R_{11}}
    \end{equation} 

In situation when $S_{22}$ cannot be measured, the only way to find $\beta_2$ is to use power balance method. This can be done by using definition (\ref{eq:betasg}). Since $R_0$ and $R_2^/$ are connected in parallel, voltage across them is the same and therefore impedance ratio  $R_0 / R_2^/$ is equal to inverse power ratio dissipated in these impedances. Assuming lossless coupler and perfectly matched $R_2$, coupling coefficient $\beta_2$ can be defined as (\ref{eq:beta_2}).   

    \begin{equation}\label{eq:beta_2}
    \beta_2 = \frac{R_0}{R_2^/} = \frac{P_{R_2^/}}{P_{R_0}} = \frac{P_{R_2}}{P_{R_0}}
    \end{equation} 
Values $P_{R_2}$ and $P_{R_0}$ can be found using S-parameters and some incident power level $P_i$ in form (\ref{eq:power lelevs}).

    \begin{equation}\label{eq:power lelevs} 
    \begin{split} 
    & P_{R_2} = P_i S_{21}^2 , \\
    & P_{R_0} = P_i - P_i S_{11}^2 - P_i S_{21}^2 . \\ 
    \end{split}
    \end{equation} 
By using (\ref{eq:power lelevs}) and (\ref{eq:beta_2}) $\beta_2$ can be extracted purely out of cavity S-parameters (\ref{eq:beta_2_from_s}).

    \begin{equation}\label{eq:beta_2_from_s}
    \beta_2 = \frac{S_{21}^2}{1 - S_{11}^2 - S_{21}^2}
    \end{equation}
Coupling $\beta_1$ can be then found from one of the equations (\ref{eq:beta_equations_pair}) in form (\ref{eq:beta_1_from_beta2}).

    \begin{equation}\label{eq:beta_1_from_beta2}
    \beta_1 = \frac{\beta_1^/}{\alpha_1}(1 + \beta_2)
    \end{equation} 
   
Since we assumed $L_2=0$, equation (\ref{eq:input_gamma1}) will reduce to $\Gamma_1 = S_{11}$. By substituting this $\Gamma_1 = S_{11}$  and $L_1$ into (\ref{eq:Gamma}) we find $\beta_1^/$ and $\alpha_1$. Coupling $\beta_1$ can be then expressed using only the S-parameters and source reflection $L_1$ in form (\ref{eq:beta_1_from_s}).       

    \begin{equation}\label{eq:beta_1_from_s}
    \beta_1 = \frac{(1+S_{11}) (1-L_1) (1-S_{11}^2)}{(1+L_1)(1-S_{11}) (1 - S_{11}^2 - S_{21}^2)}
    \end{equation} 
Finally, $Q_0$ can be found then in the form (\ref{eq:Q_0_power_balance}).   

    \begin{equation}\label{eq:Q_0_power_balance}
    Q_0 = Q_L \Bigg | \frac{2(1+S_{11})(1-L_1 S_{11}) }{(1+L_1)(1-S_{11}^2-S_{21}^2)}  \Bigg |
    \end{equation}

By continue this reasoning for N-port cavity one can show fairness of expression (\ref{eq:beta_1_for_n_port}) for $\beta_1$ which follows from the equation (\ref{eq:quation_summ}); and show fairness of expression (\ref{eq:beta_n_for_n_port}) for $n>1$ which follows from the power balance.    

    \begin{equation}\label{eq:beta_1_for_n_port}
    \beta_1 = \frac{\beta_1^/}{\alpha_1} \Bigg (1 + \sum_{n=2}^N \beta_n \Bigg )
    \end{equation}
    
    \begin{equation}\label{eq:beta_n_for_n_port}
    \beta_n = \frac{S_{n,1}^2}{1- \sum_{i=1}^N S_{i,1}^2}, \ (n>1)
    \end{equation}
Therefore for multiport cavity with arbitrary cavity and test ports coupling $Q_0$ can be extracted purely from the S-parameters (\ref{eq:Q_0_power_balance_Nport}).  
    \begin{equation}\label{eq:Q_0_power_balance_Nport}
    \begin{split} 
    & Q_0 = Q_L \Bigg |\frac{(L_1-1) (S_{11}+1)}{(L_1+1)(S_{11} - 1)} \Sigma_1 + \Sigma_1  \Bigg | , \\
    &\text{where:} \\
    & \Sigma_1 = 1 + \sum_{n=2}^N \frac{S_{n,1}^2}{1-\Sigma_2}, \quad \Sigma_2 = \sum_{i=1}^N S_{i,1}^2 .\\
    \end{split}
    \end{equation}

\subsection{Second order approximation}
When not only load match but also the RF source match are assumed perfect $L_1=0, \ \alpha_1 = 1$ equations (\ref{eq:cav_s11_s21}), (\ref{eq:beta_1_for_n_port}), (\ref{eq:Q_0_power_balance_Nport}) reduce to the second order approximation (\ref{eq:cav_s11_s21_secnd_order}), (\ref{eq:beta_1_for_n_port_secnd_order}), (\ref{eq:Q_0_power_balance_Nport_secnd_order}). 
      
    \begin{equation}\label{eq:cav_s11_s21_secnd_order} 
    \begin{split}
    S_{11} = R_{11}, \
    S_{21} = T_{21}.
    \end{split}
    \end{equation} 

    \begin{equation}\label{eq:beta_1_for_n_port_secnd_order}
    \beta_1 = \beta_1^/ (1 + \beta_2 +\hdots+ \beta_n )
    \end{equation}    

    \begin{equation}\label{eq:Q_0_power_balance_Nport_secnd_order}
    Q_0 = Q_L \Bigg | \frac{(S_{11}+1)}{(1-S_{11} )} \Sigma_1 + \Sigma_1  \Bigg | 
    \end{equation}    

Because of its simplicity, many SRF facilities are using second order approximation, which unfortunately is leading to significant errors if the cavity is not critically coupled ($S_{11} \neq 0$). It is practically hard to make $L_1 = 0$, which is usually port reflection of high-power circulator and can change significantly with temperature. Such measurements are typically done with  power meters and lack of information about the phase of $S_{11}$ requires introducing additional coefficient $C_\beta = \pm 1$ before $S_{11}$ in (\ref{eq:Q_0_power_balance_Nport_secnd_order}) to determine if the cavity is undercoupled or overcoupled.

\section{$Q_0$ Accuracy Assessment}

Schappert \cite{Holzbauer} have pointed out that even in the critically coupled cavity, error in the loaded quality factor measurement due to source mismatch can reach 20\% and higher. Our calculations show that in fact loaded quality factor error does not depend on mismatches but coupling coefficients $\beta_n$ contain such mismatch errors, which eventually lead to errors in $Q_0$ determination. Study of $\beta_n$  errors therefore presents practical interest.   

For exact solution (\ref{eq:q0_sparam}) accuracy is mostly determined by the accuracy of S-parameters \footnote{For highly reflective devices such as overcoupled cavities or cryomodules, high accuracy of S-parameter measurement can be enabled by special methods, see \cite{impedance1}, \cite{impedance2}. } and reflection coefficients. This accuracy depend on the specific test stand or VNA and calibration kit being used and should be evaluated for each measurement using special metrological techniques (see p.178 in \cite{dunsmore}). Such measurements using (\ref{eq:q0_sparam}) provide ultimate accuracy limited by the VNA and results are traceable to national standards. Next we will we show how approximations deviate from the results obtained with (\ref{eq:q0_sparam}).

\subsection{First order approximation}
\begin{figure}[!htb]
\includegraphics[width=0.4\textwidth]{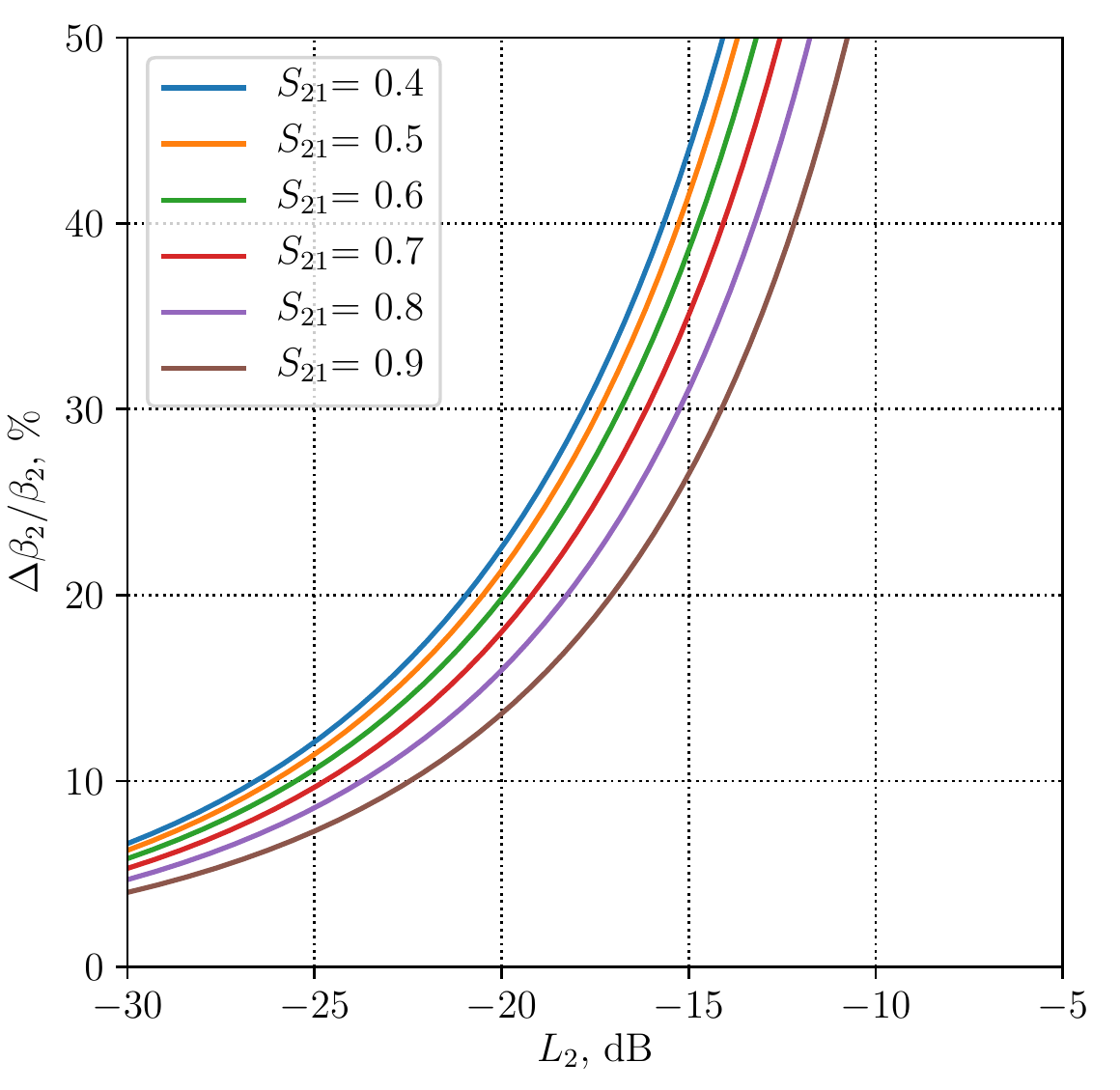}
\caption{\label{fig:beta_2_err} First order approximation error of $\beta_2$ measurement vs load $L_2$ reflection with various values of transmission coefficient $S_{21}$.}
\end{figure}

\begin{figure}[!htb]
\includegraphics[width=0.4\textwidth]{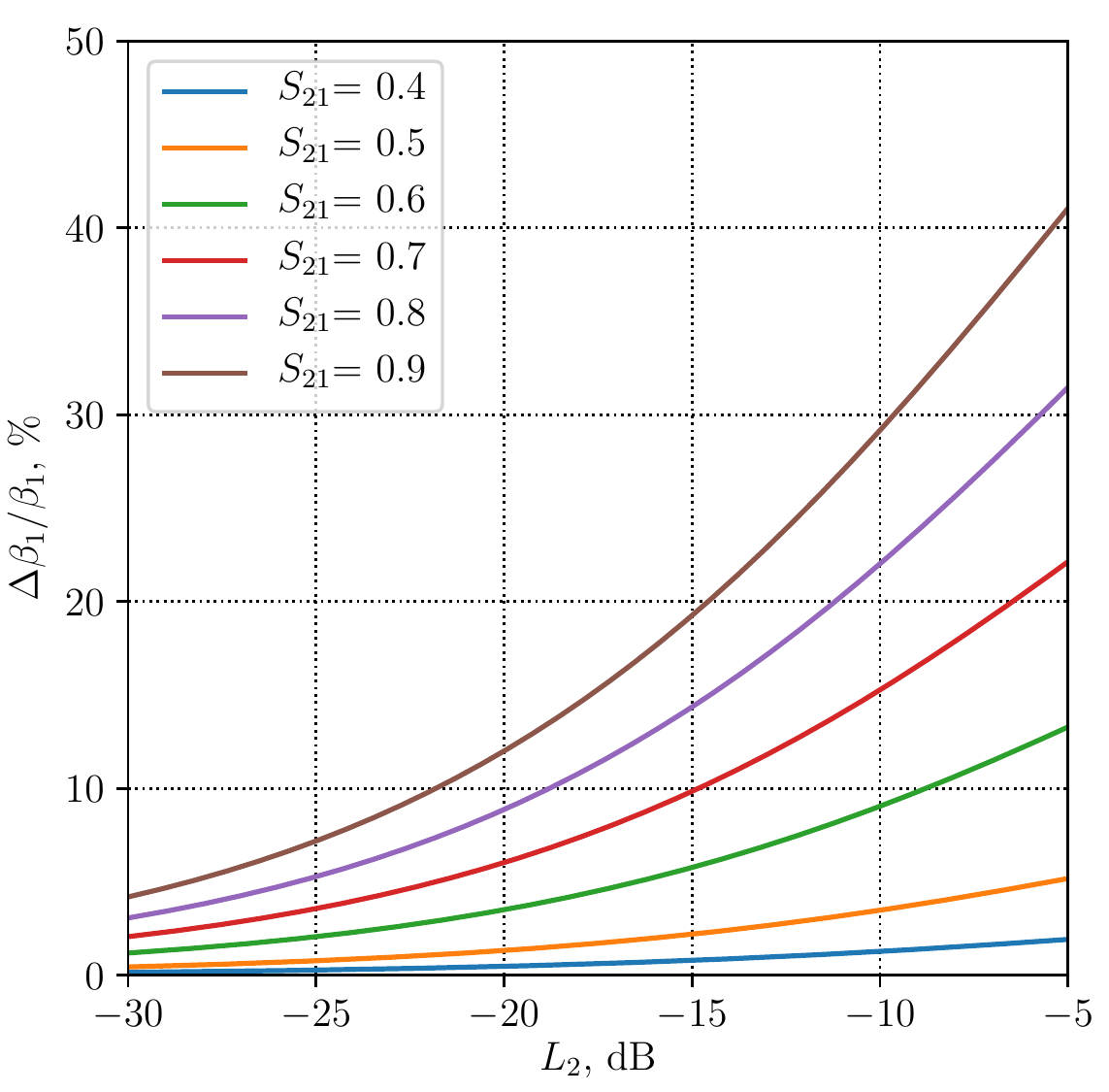}
\caption{\label{fig:beta_1_err} First order approximation error of $\beta_1$ measurement vs load $L_2$ reflection with various values of transmission coefficient $S_{21}$.}
\end{figure}

The first order approximation equations (\ref{eq:reflection_transmission}) demand for $L_2 \to 0$, this assumption allows to extract value of $\beta_2$ using power balance method. Therefore, reflection coefficient $L_2$ influences on the errors of $\beta_2$ and $\beta_1$ measurement that \textit{cannot} be corrected. These errors can be analytically evaluated using the exact solution (\ref{eq:q0_sparam}) for $\beta_2$ and $\beta_1$  in the form of expressions (\ref{eq:beta_2_errorr})  that depend only on $S_{xx}$ and $L_1, L_2$; fractional error of $\beta_2$  with typical parameters of critically coupled two-port cavity $S_{11}=0.01$ ($\beta_1 = 1.02$), $S_{22}=-0.818$ ($\beta_2=0.1$) is evaluated in fig.~\ref{fig:beta_2_err} and the fractional error of $\beta_1$ it is evaluated in fig.~\ref{fig:beta_1_err}.

    \begin{equation}\label{eq:beta_2_errorr}
    \begin{split}
    & \Delta \beta_2 =  \beta_2 \Big \rvert_{L_2=0}  - \beta_2 \Big \rvert_{L_2 \neq 0}, \\    
    & \Delta \beta_1 =  \beta_1 \Big \rvert_{L_2=0}  - \beta_1 \Big \rvert_{L_2 \neq 0}.
    \end{split}
    \end{equation}

\begin{figure}[!htb]
\includegraphics[width=0.4\textwidth]{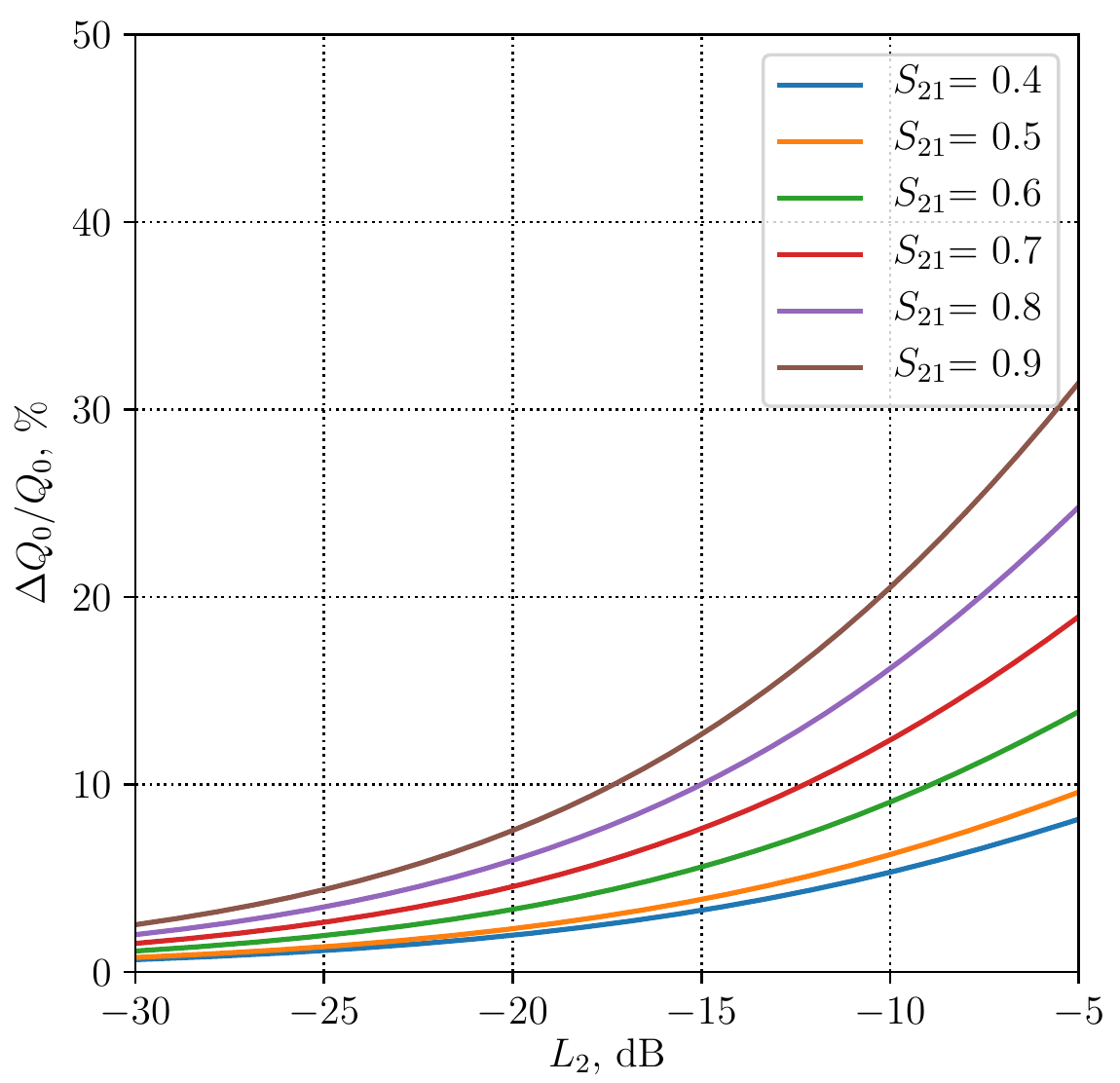}
\caption{\label{fig:q0_err} First order approximation error of $Q_0$ measurement of critically coupled cavity vs load $L_2$ reflection with various values of transmission coefficient $S_{21}$.}
\end{figure}

Assuming the error of $Q_L$ measurement is negligible, using definition (\ref{eq:Q0}), the fractional error of $Q_0$ measurement for two-port cavity can be found in the form (\ref{eq:q_0_error}). Behavior of this error is shown in fig.~\ref{fig:q0_err}.

    \begin{equation}\label{eq:q_0_error}
    \frac{\Delta Q_0}{Q_0} =  \frac{  \Delta \beta_1 + \Delta \beta_2}{1+\beta_1+\beta_2}    
    \end{equation} 

It is practically very difficult to reach $L_2 = -30$~dB, in fact in most cases $-25$~dB is the best reflection one can achieve in a broadband load used for vertical tests of SRF cavities. Therefore, fig.~\ref{fig:beta_1_err} and fig.~\ref{fig:beta_2_err} lead to important result: for critically coupled cavity and matched source $L_1=0$ practically achievable accuracy in power balance method is around $8 \hdots 12 \%$ for $\beta_2$ and $4 \hdots 8 \%$ for $\beta_1$ (since $S_{21}<0.5$ typically), so $Q_0$ error cannot be better than 5 \% using first order approximation for near critically coupled cavity ($\beta_1 = 1.02$), for the wider range of couplings accuracy will depend on the residual errors of the test fixture.

\subsection{Second order approximation}

For the second order approximation reflection coefficients $L_2$ and $L_1$ influence the errors of $\beta_2$ and $\beta_1$ measurement that cannot be corrected. However, $L_2$ can be made relatively small, so we study influence of $L_1$ on the error. 

\begin{figure}[!htb]
\includegraphics[width=0.4\textwidth]{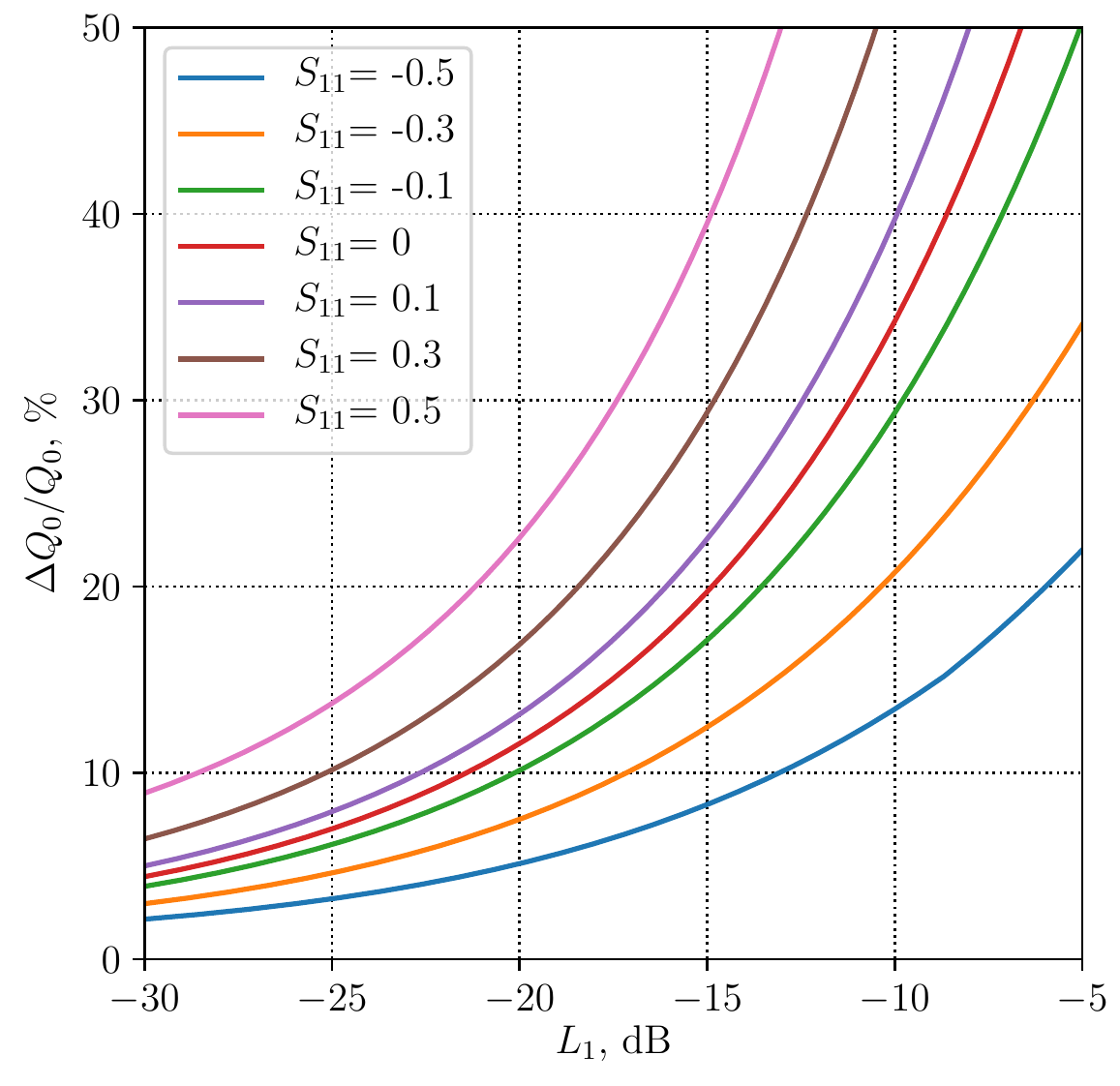}
\caption{\label{fig:q0_errr} Second order approximation error of $Q_0$ measurement cavity with $S_{21}=0.4$, $L_2 = -25$~dB and various input couplings $S_{11}$ vs load $L_1$ reflection.}
\end{figure}

Using equations (\ref{eq:q0_sparam}) and (\ref{eq:beta_2_error}) we evaluate $Q_0$ accuracy for the case when cavity is undercoupled or overcoupled, and RF source has some mismatch ($L_1 \neq 0$). Based on the results from the previous section we will fix $L_2=-25$~dB, $S_{21}=S_{12}=-8$~dB, $S_{22}=-1.7$~dB. These values give the following result shown in fig.~\ref{fig:q0_errr}. 

In can be seen that \textit{even with critical coupling} ($S_{11}=0$) of the cavity input port, error of $Q_0$ measurement exceeds 10 \%, which is higher than previous estimations because \cite{Melnychuk} assumed $L_1=L_2=0$ and \cite{Holzbauer} assumed $L_2=0$. In fact, second order approximation assumption that $L_1 = 0 $ and $L_2=0$ in majority cases is not valid. When $L_2$ and especially $L_1$ differ from zero, error depends on the type of cavity coupling (undercoupled for negative $S_{11}$, overcoupled for positive $S_{21}$). The behavior when for undercoupled mode error is lower than for overcoupled is in agreement with the previous results \cite{Holzbauer}. 

\begin{figure}[!htb]
\includegraphics[width=0.4\textwidth]{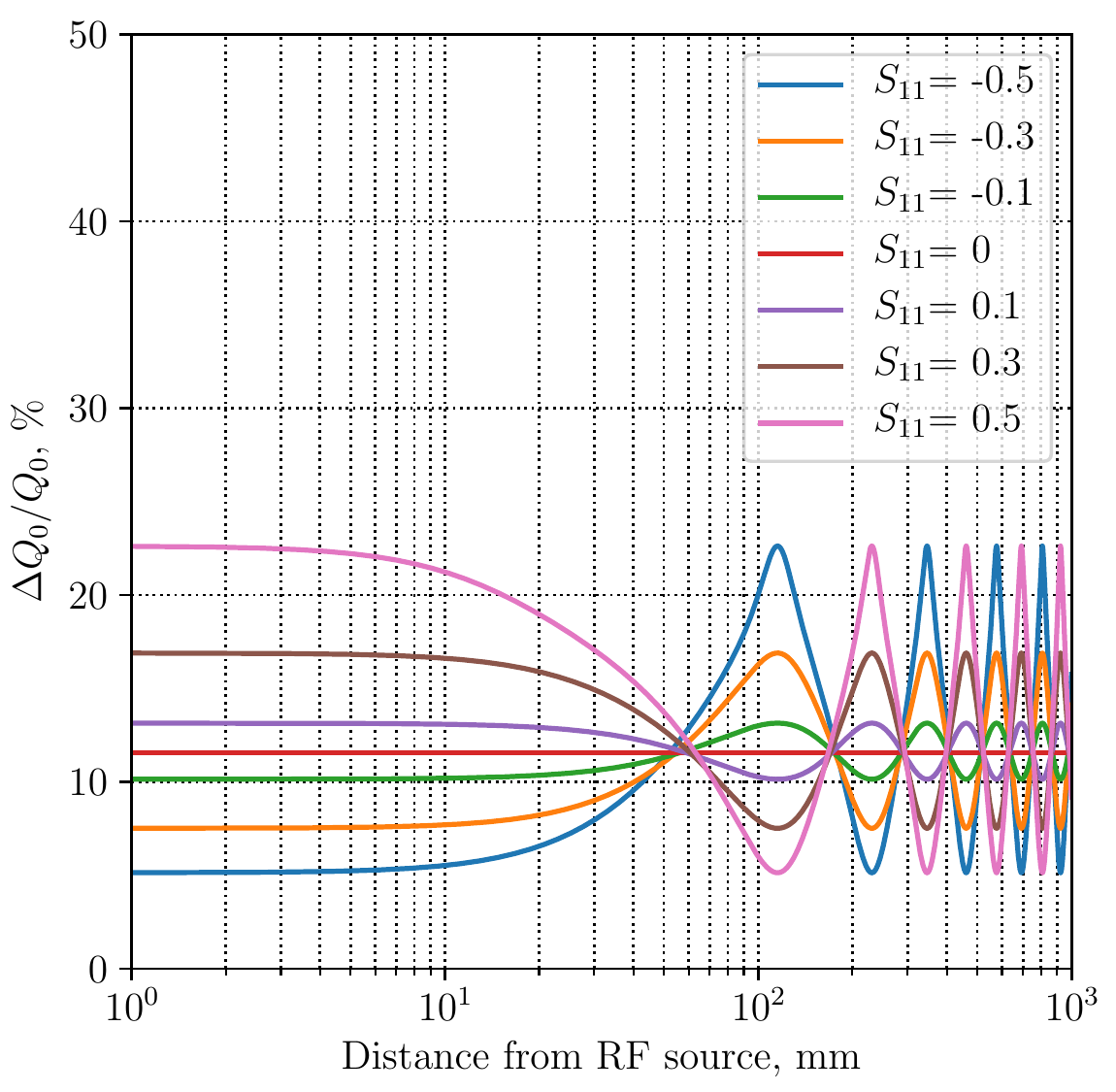}
\caption{\label{fig:q0_err_vector} Second order scalar approximation error of $Q_0$ measurement cavity with $S_{21}=0.4$, $L_2 = -25$~dB, $L_1 = -20$~dB,  and various input couplings $S_{11}$ vs distance from the RF source.}
\end{figure}

However, coupling seen by the test port (output of the directional coupler in the test setup) depends on the phase of the reflected signal and without vector calibration does not depend on type of cavity coupling. Therefore, if the cavity is not critically coupled, $Q_0$ error will depend on distance between test port (RF reflectometer) and cavity input port. This is demonstrated in fig.~\ref{fig:q0_err_vector} where $S_{11}$ is multiplied by $\exp(j 2\pi l /\lambda )$, where $l$ is distance and $\lambda$ is wavelength at 1.3 GHz. This vector error cannot be corrected with scalar test setup based on powermeters, a full featured vector receiver should be used instead.    

    \begin{equation}\label{eq:beta_2_error}
    \begin{split}
    & \Delta \beta_2 =  \beta_2 \Big \rvert_{L_1 = L_2 =0}  - \beta_2 \Big \rvert_{L_2 = -25 \text{\ dB},\  L_1 \neq 0}, \\    
    & \Delta \beta_1 =  \beta_1 \Big \rvert_{L_1 = L_2 =0}  - \beta_1 \Big \rvert_{L_2 = -25 \text{\ dB},\  L_1 \neq 0} .
    \end{split}
    \end{equation}    

\section{Verification by Simulation}
\begin{figure*}[!htb]
\includegraphics[width=1\textwidth]{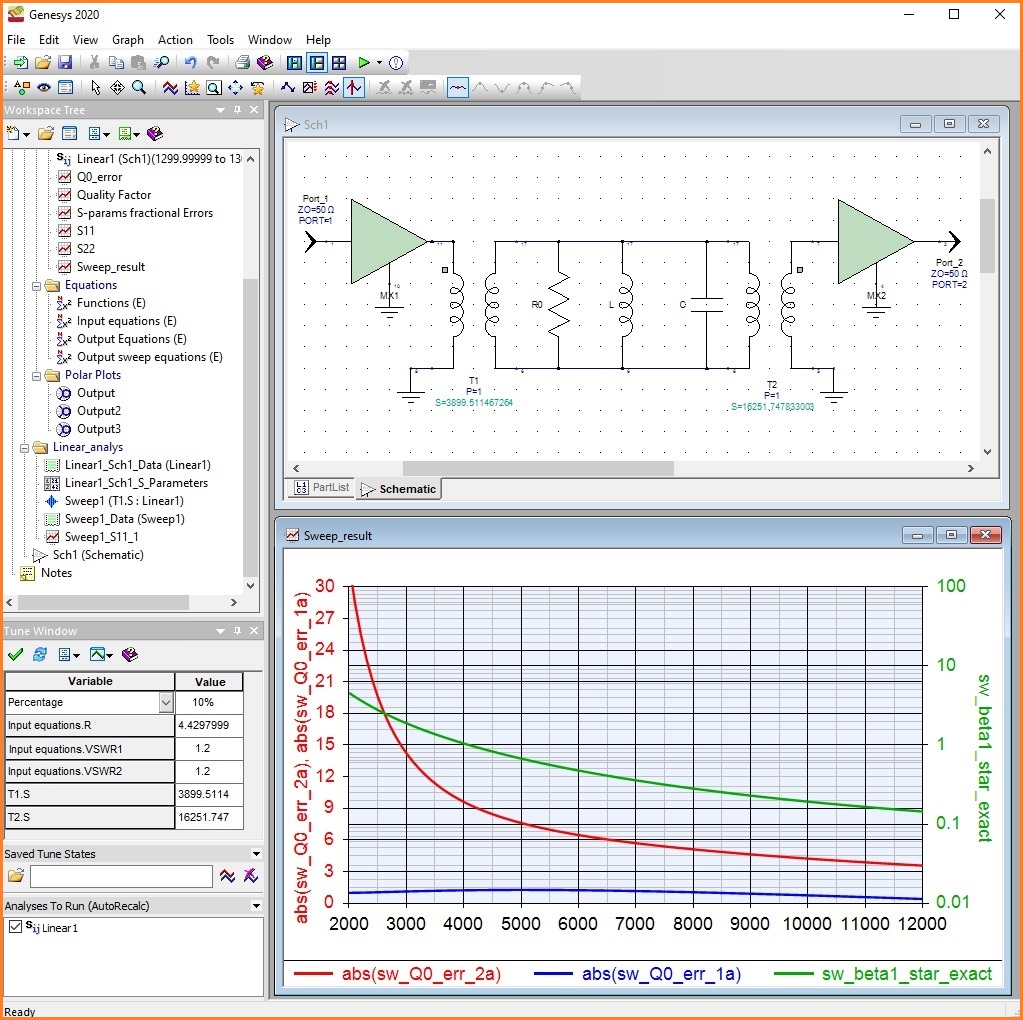}
\caption{\label{fig:genesys} Simulation of the cavity equivalent circuit in Genesys 2020. Cavity couplers (transformers) T1 and T2 are connected to Port1 and Port2 via gain blocks MX1 and MX2. Tuning VSWR1 and VSWR2 of these gain blocks (see Tune Window in the left bottom corner) allows to simulate mismatches of the test fixture. Tuning number of turns S in the secondaries of transformers T1 and T2 allows to simulate various coupling of the cavity ports. Bottom plot shows percentage of absolute systematic error in $Q_0$ measurement (left vertical scale) for the 1st order approximation "sw\_Q0\_err\_1a" and for the 2nd order approximation "sw\_Q0\_err\_2a" vs the T1 transformer secondary winding number of turns on the horizontal scale. Equivalent input cavity coupling corresponding to this number of turns is shown on the right vertical scale "sw\_beta1\_star\_exact".  Transformer T2 is tuned so that cavity output is weakly coupled $\beta_2 < 0.1$ to Port2.}
\end{figure*}

We used Keysight Genesys 2020 RF/microwave circuit synthesis and simulation software to verify the obtained results. The simulator window is shown in fig.~\ref{fig:genesys}. Multiple scenarios can be simulated using this model. For us it was very interesting to compare the simulation with the analytic result in fig.~\ref{fig:q0_errr}. It can be seen that simulation is in good agreement: even with perfect input match of the cavity ($S_{11}$=0 in fig.~\ref{fig:q0_errr}, sw\_beta1\_star\_exact=1 in fig.~\ref{fig:genesys}), error of the second order approximation "sw\_Q0\_err\_2a" is close to 10\% for VSWR = 1.2 of the test ports  (-21 dB), while first order approximation "sw\_Q0\_err\_1a" is below 5 \%. Multiple tests of the model in fig.~\ref{fig:q0_errr} indicated correctness of the obtained expressions (\ref{eq:quation_summ_gen_solution}), (\ref{eq:q0_sparam}), (\ref{eq:Q_0_power_balance_Nport}) and (\ref{eq:Q_0_power_balance_Nport_secnd_order}).   

\section{Conclusion}

In this paper we analyzed features of cavity intrinsic quality factor measurements. We derived general formula for the cavity coupling coefficients with any number of ports based on the lumped-element model. Using this result, we demonstrated how to extract cavity intrinsic quality factor only from S-parameters in form of the exact solutions and two approximations. We showed that most SRF facilities that utilize scalar error correction for cavity measurements are in fact using the second order approximation that gives error order of 10\% even for critically coupled cavities. We found this error larger than in previous reports, since we have also included mismatches of the output port, that have not been considered before. Finally, we demonstrated correctness of the obtained formulas in two ways: by direct analytic comparison and by using simulation software. The ultimate $Q_0$ accuracy provided by exact solution cannot be achieved for superconducting accelerating cavity because of technical limitations, however even for overcoupled or undercoupled cavities relatively low error $\approx$~5\% can be achieved by using the proposed first order approximation.

Practical implementation of the obtained expressions (\ref{eq:q0_sparam}) and (\ref{eq:Q_0_power_balance_Nport}) requires replacement of scalar RF system based on power meters with full-featured vector receivers and vector calibration system. Such RF system and its calibration are currently under development at Fermilab and will be subject for further studies.

\begin{acknowledgements}
Author would like to acknowledge R. Pilipenko for help in understanding features of existing VTS system at Fermilab and A. Romanenko for the support of this work, as well as A. Babenko for discussion of microwave measurements and error models. This manuscript has been authored by Fermi Research Alliance, LLC under Contract No. DE-AC02-07CH11359 with the U.S. Department of Energy, Office of Science, Office of High Energy Physics.
The data that support the findings of this study are available from the corresponding author upon reasonable request.
\end{acknowledgements}

\end{document}

%% file: preamble.tex
\usepackage{amsthm}
\usepackage{mathtools}
\usepackage{physics}
\usepackage{xcolor}
\usepackage{graphicx}
\usepackage[left=23mm,right=13mm,top=35mm,columnsep=15pt]{geometry} 
\usepackage{adjustbox}
\usepackage{placeins}
\usepackage[T1]{fontenc}
\usepackage{lipsum}
\usepackage{csquotes}